\documentclass{article} 
\usepackage{graphicx}

\usepackage[scaled]{helvet}  
\usepackage{courier} 
\usepackage{eulervm} 
\usepackage[T1]{fontenc}
\usepackage{color}
\usepackage{pict2e}
\usepackage{mathrsfs}
\usepackage{hyperref}
\usepackage{amssymb,amsmath,amsthm}
\usepackage{bbold}
\usepackage[cal=boondox]{mathalfa}

\normalfont

\newtheorem{Thm}{Theorem}[section]
\newtheorem{theorem}[Thm]{Theorem}
\newtheorem{proposition}[Thm]{Proposition}

\newtheorem{remark}[Thm]{Remark} 
 
\newtheorem{definition}[Thm]{Definition}

\begin{document}

\title{The Pricing of Quanto Options \\
\quad \\
\large An empirical copula approach 
}

\author{Rafael Felipe Carmargo 
Prudencio\footnote{rafael.camargo.prudencio@gmail.com, 
Department of Applied Mathematics, University of 
S\~ao Paulo (USP), Brazil.}
\\
and Christian D.\ J\"akel\footnote{jaekel@ime.usp.br, 
Department of Applied Mathematics, University of 
S\~ao Paulo (USP), Brazil.} }        

\maketitle

\begin{abstract}
The quanto option is a cross-currency derivative in which the 
pay-off is given in foreign currency and then converted to 
domestic currency, through a constant exchange rate, 
used for the conversion and determined
at contract inception. Hence, the dependence 
relation between the option underlying asset price and the 
exchange rate plays an important role in quanto option pricing.  

In this work, we suggest to use \emph{empirical} copulas to price quanto options. 
Numerical illustrations show that the flexibility provided by this 
approach, concerning the dependence relation of the two underlying 
stochastic processes, results in non-negligible pricing differences when 
contrasted to other models.
\end{abstract}

\tableofcontents

\section{Introduction}

The \emph{quanto option} is a cross-currency contract. The payoff 
is defined with respect to an underlying asset or index in one currency, 
but for payment, the payoff is converted to another currency.  The 
constant exchange rate is established at contract inception. 
Hence, the modelling of the dependence relation between the underlying 
asset and the exchange rate (which are both market observable 
variables), is mandatory for quanto options pricing. In this work, 
we propose  a new approach, based on empirical copula, 
to price quanto options. We compare this approach with 
what is hereafter named the practitioners model (based 
on the Black-Scholes framework) 
and the Dimitroff-Szimayer-Wagner (DSW)
framework \cite{DSW}. Without loss of generality, 
only call options are analysed, 
with the dividend yield of the underlying asset set to zero.

The practitioners' approach is based on the assumptions that ``asset prices 
follow a geometric Brownian motion'' and ``volatility is constant''. 
Stochastic volatility models, such as the one 
proposed in \cite{DSW}, relax 
the ``volatility is constant'' assumption. 
In the quanto option context, the dependences 
among the relevant variables can considerably 
impact the pricing. Both the practitioners' approach and 
the DSW model \cite{DSW} use a 
constant correlation in order to address this issue. However, 
financial quantities (including the underlying asset and 
the exchange rate) can be related in a non-linear way 
(see, \emph{e.g.}, Teng et al.~\cite{TEG}). 
Hence a simple constant correlation cannot 
\emph{fully} 
represent the dependence relation between the relevant  
variables. 

The copulas framework, which we propose, 
intends to provide a more flexible framework to set 
the dependence relation between the market 
variables used in the pricing of quanto options. 
Besides, the empirical copula model 
(just like the DSW model) can adapt 
to a non-constant volatility smile. 
Before we start our discussion, we would like to note
that we are aware of the shortcomings of our approach: it 
is computationally expensive and does not offer analytical tractability.

\section{The quanto process}

A \emph{quanto call option} is a financial instrument that gives the 
holder the right, but not the obligation, to buy an \emph{underlying asset} $S_f$, 
\emph{quoted in a foreign currency} (FOR), at a predetermined price $K$ (given in 
units of FOR currency), at maturity time $T$. The payoff amount, if 
positive, is \emph{converted to the domestic currency} (DOM) at an 
exchange rate $q ( \equiv \frac{DOM}{FOR})$. The latter
is predetermined at the contract inception.
Hence, the payoff, at maturity time $T$, is 
	\begin{equation}
	\label{(1)}
		C_q ( T ) = \max \bigl\{ q(S_f (T ) - K), 0 \bigr\} \; .
	\end{equation}

\subsection{The practitioners approach}

From the risk-neutral pricing formula 
it follows that the price $c_q$  
of a quanto call option at time $t = 0$ is  
	\begin{equation}
	\label{(2)}
		c_q(0) = {\rm e}^{ - rT} \; \mathbb{E}_{\mathbb{Q}}
		\Bigl[ \max \{ q ( S_f (T) - K ) , 0 \} \Bigr] \; ,   
	\end{equation}
where $\mathbb{Q}$ is the \emph{domestic risk-neutral measure}
and $\mathbb{E}_\mathbb{Q}$ denotes the associated expectation value. 

We now derive the stochastic differential equation for
$S_f ( T )$  under $\mathbb{Q}$. We assume that, under the 
domestic risk-neutral measure, 
	\begin{equation}
	\label{(3)}
		{\rm d} S_f (t) = \mu_{S_f} {\rm d} t +\sqrt{ V_1} \, S(t) \, 
		{\rm d} W_{\mathbb{Q}_1} (t) \; ,
	\end{equation}
where $\mu_{S_f}$ is the (unknown)  
\emph{drift} of $S_f (t)$ and $W_{\mathbb{Q}_1}$ represents 
a Brownian motion. The volatility is denoted by $\sqrt{ V_1}$.

The stochastic process of the exchange rate $Q(t) ( \equiv \frac{DOM}{FOR} )$ 
under the domestic risk neutral measure is 
	\begin{equation}
	\label{(4)}
		{\rm d} Q(t) = Q(t) \Bigl[ (r - r_f) {\rm d} t 
		+ \sqrt{V_2} \; {\rm d} W_{\mathbb{Q}_2} (t) \Bigr] \; ,
	\end{equation}
with
	\begin{equation}
	\label{(5)}
		W_{\mathbb{Q}_2} (t) = \rho(S_f, Q) W_{\mathbb{Q}_1} (t) 
		+ \sqrt{1 - \rho(S_f, Q)^2} \, \;  W_{\mathbb{Q}_3} (t) \; ,
	\end{equation}
a second Brownian motion, 
correlated with the Brownian motion $W_{\mathbb{Q}_1}$. On the other hand, 
$W_{\mathbb{Q}_3}$ is a Brownian motion, which is independent
from $W_{\mathbb{Q}_1}$. As can be read off from 
\eqref{(5)}, the \emph{infinitesimal correlation between 
the increments of $S_f$ and $Q$} is denoted by $\rho(S_f, Q) $. 

In order to derive the drift $\mu_{S_f}$, we express $S_f(t)$ in the domestic currency:  
we multiply  $S_f (t)$ by $Q(t)$, setting
	\[
		S_d (t) \doteq Q(t) S_f (t) \; .
	\]
From It\o's product rule it now follows that
 	\begin{align*}
		{\rm d} \bigl( S_d (t) \bigr) 
		& = Q(t) S_f (t) \left[ \; \mu_{S_f} +r - r_f + \rho(S_f, Q) \sqrt{V_1 V_2} \; \right] 
		{\rm d} t 
		\\
		& \qquad \qquad \qquad \qquad \qquad
		+ \sqrt{V_1} \, {\rm d} W_{\mathbb{Q}_1} (t) 
		+ \sqrt{V_2} \, {\rm d} W_{\mathbb{Q}_2} (t) \; .
	\end{align*}
Under the domestic risk neutral measure, the drift of $S_d(t)$ is equal to $r$. 
Thus, it follows that
	\begin{align*}
		\mu_{S_f} & = r - \left[ \, r - r_f + \rho(S_f, Q) \sqrt{V_1 V_2} \; \right] \; . 
	\end{align*}
Inserting this expression into \eqref{(3)}, we find 
	\[
		{\rm d} S_f (t) = r - \left[ r - r_f + \rho(S_f, Q) \sqrt{V_1 V_2} \; \right]  {\rm d} t 
		+ \sqrt{V_1} \, S(t) \, {\rm d} W_{\mathbb{Q}_1}(t) \; .
	\]
Since we know the dynamics of $S_f (t)$, we are now able 
to compute the expectation \eqref{(2)}. In fact, the diffusion of 
$S_f$ is of the same form as the diffusion process 
for a dividend paying stock, with dividend rate 
	\[
	 	q = r - r_f + \rho ( S_f, Q) \sqrt{V_1 V_2} \; . 
	\]
Whence, the computation of expectation \eqref{(2)} 
gives the price of the vanilla call option on a dividend-paying stock: 
	\[
		c^q (0) = q \cdot BS \left( S_f (0) 
		{\rm e}^{ - ( r - r_f + \rho(S_f, Q) \sqrt{V_1 V_2})
		T }, K , \sqrt{V_1} , T , r \right) \; .
	\]
Here $BS (a, b, c, d, e)$ stands for the traditional Black-Scholes 
formula, with $a$ the underlying asset spot price,  $b$ the strike value,  
$c$ the volatility, $d$ the time to maturity, and $e$ the risk-free interest rate.
\color{black}

The final step in the practitioners approach is to replace the 
constant volatilities,  $\sqrt{V_1}$ and $\sqrt{V_2}$, by \emph{at the money} 
or \emph{at the strike values}:
	\begin{equation}
	\label{(6)} 
	c^q_p (0) = q \cdot BS \left( S_f (0) {\rm e}^{ - T \left( r - r_f +
	\rho(S_f, Q) \sqrt{V_{1}^{atm} V_{2}^{atm}} \,\right) }, K, 
	\sqrt{V_{1}^{strike}} , T , r 
	\right)
	\; .
	\end{equation}
Equation \eqref{(6)} is the $V^d_{black}$ approximation from Le Floc'h \cite{F}; 
in fact, it is one of the three approximations  
studied within \cite{F}. Note that $V^{atm}_i$, $i=1,2$, 
in $\rho(S_f, Q) \sqrt{V^{atm}_1 V^{atm}_2}$, must be the 
at-the-money value (not the at the strike value $V^{strike}_i$, $i=1,2$), as
otherwise the price of the quanto forward contract would depend on 
the option strike (an exogenous factor).

\subsection{The Dimitroff-Szimayer-Wagner (DSW) framework}
\label{sec:2.2} 

The DSW approach consists in the use of the following 
diffusion processes to simulate values of $S_f(t)$ (named $S(T)$ 
in their work) 
and $Q^{-1}(T)$ (named $C(T)$ in their work), in order to compute 
expectation \eqref{(7)} below and to obtain the quanto option price value: 
	\begin{align*}
	\begin{pmatrix}
	{\rm d} S_f (t) \\
	{\rm d} V_1(t) \\
	{\rm d} Q^{-1}(t) \\
	{\rm d} V_2(t) 
	\end{pmatrix}
	& = \begin{pmatrix}
	( r_f(t)  -  d(t) ) S_f (t) \\
	\kappa_1 ( \overline{V_1} - V_1(t))  \\
	( r_f(t) - r(t)) Q^{-1}(t) \\
	\kappa_2 ( \overline{V_2} - V_2(t))  
	\end{pmatrix} {\rm d}t 
	\\
	& \quad + \begin{pmatrix}
	\sqrt{V_1(t)} \, S_f (t) & 0 & 0 & 0 \\
	0 & \eta_1 \sqrt{V_1(t)} & 0 & 0 \\
	0 & 0 & \sqrt{V_2(t)} \, Q^{-1}(t) & 0  \\
	0 & 0 & 0 & \eta_2 \sqrt{V_2(t)}  
	\end{pmatrix}
	\\
	& \qquad \times \begin{pmatrix}
	1 & 0 & 0 & 0 \\
	\rho_1 & \sqrt{1- {\rho_1}^2} & 0 & 0 \\
	\rho & 0 & \sqrt{1- \rho^2} & 0  \\
	\rho \rho_2 & 0 & \rho_2 \sqrt{1- \rho^2} & \sqrt{1- {\rho_2}^2}  
	\end{pmatrix}
	\begin{pmatrix}
	\overline{{\rm d}W_1} (t) 
	\\
	\overline{{\rm d}W_2} (t) 
	\\
	\overline{{\rm d}W_3} (t) 
	\\
	\overline{{\rm d}W_4} (t) 
	\end{pmatrix}
	\end{align*}
where $(S_f (t), V_1(t))$ models the stock price and its variance, 
and $(Q^{-1}(t), V_2 (t))$ the foreign exchange 
rate and its variance with correlation 
$\rho_1$ and $\rho_2$, respectively. 
The correlation between the Brownian 
motions of the $S_f(t)$ and $Q^{-1}(t)$ diffusions is denoted 
by $\rho \equiv \rho(S_f, Q^{-1})$. 
The domestic risk-free interest rate 
is denoted by $r(t)$, the foreign risk free interest 
rate by $r_f(t)$, and the 
continuous dividend yield of the stock by $d(t)$. 
As the Heston model is one of the main building blocks 
of the DSW approach, the constants $\overline{V_i}$, 
$\kappa_i$ and $\eta_i$ have the 
traditional meaning, 
\emph{i.e.}, 
$\overline{V_i}$ is the long run variance, 
$\kappa_i$ is the rate at which $V_i(t)$ 
reverts to $\overline{V_i}$, and~$\eta_i$
determines the variance of the process $V_i(t)$, $i=1,2$. 

Besides, it is necessary to set $V_i(0)$, which is the initial variance, 
in order to get the full representation of the DSW approach 
in the risk-neutral format. Finally, the parameters in the 
equations above can be compiled in the Heston vector 
of parameters $\varphi_{S_f}$ and $\varphi_{Q^{-1}}$, with
	\[
		 \varphi_{S_f} = \bigl( \rho_1, \kappa_1, \overline{V_1}, V_1(0), \eta_1 \bigr)
		 \quad \text{and} \quad
		 \varphi_{Q^{-1}} = \bigl( \rho_2, \kappa_2, \overline{V_2}, V_2(0), \eta_2 \bigr)
		 \; . 
	\]
These Heston vectors of parameters are calibrated 
with market data in order to take into account the 
respective market volatility smiles.

\subsection{Risk neutral pricing from a foreign investor's perspective}

Our new framework (as well as  
the DSW model) bases the quanto option pricing 
on the diffusion processes of $S_f (t)$ and $Q^{-1} (t)$, 
$0 \le t \le T$, under the 
\emph{foreign risk neutral measure} $\mathbb{Q}_f$.   
From a foreign investor's  
perspective, the payoff, given in FOR currency, is
	\[
		C^f_q (T) =   Q^{-1}(T) \max \bigl\{ q ( S_f(T) - K),0 \bigr\} \; .
	\]
Here $Q^{-1} (t) ( \equiv \frac{FOR}{DOM})$
is the exchange rate quoted as foreign currency per unit of domestic 
currency. 

\goodbreak
From the risk neutral pricing formula, the option value (in FOR currency) is given by
	\[
		c_q^f  (0) = {\rm e}^{-r_f T} \, \mathbb{E}_{\mathbb{Q}_f} \; 
		\Bigl[ Q^{- 1} (T)
		\max \bigl\{ q ( S_f(T) - K),0 \bigr\} \Bigr]\; .
	\]
A non-arbitrage argument can be used to value the option in DOM currency:
	\begin{equation}
	\label{(7)}
	c_q (0) = Q( 0 ) \, {\rm e}^{-r_f T}
	\mathbb{E}_{\mathbb{Q}_f} \Bigl[
		Q^{-1}(T ) \max \bigl\{ q(S_f (T ) - K ), 0 \bigr\} \Bigr] \; . 
	\end{equation}
Equation \eqref{(7)} sets a starting point for quanto option pricing. 

\section{The quanto option pricing under the empirical copula approach}

A variety of methodologies can be used in 
order to compute the expectation in \eqref{(7)}. We like 
to make the pricing of quanto options as adaptable as possible to 
the dependence relation between $S_f(T)$ 
and $Q^{-1}(T)$. At the same time, our
approach is capable to adapt to the market volatility smiles.

The expectation in equation \eqref{(7)} involves two random 
variables, namely $S_f (T )$ and $Q^{-1}(T)$, hence, one approach to solve it, 
is to estimate the bi-variate \emph{cumulative distribution 
function} (CDF) of these random variables, under the probability
measure $\mathbb{Q}_f$, and to compute the expectation based 
on simulations of this CDF. We denote the CDF by $H(s_f (T), q^{-1}(T))$
in this text, where $s_f (T)$ and $q^{-1}(T)$ are the possible outcomes of the 
random variables $S_f (T )$ and $Q^{-1}(T)$, respectively. 
The main ingredient in our analysis is Sklar's Theorem. 
It ensures the existence of a \emph{copula}, \emph{i.e.}, 
a function $C \colon [0,1]^d \to \mathbb{R}^+$ with the following properties~\cite{FS}: 
\begin{itemize}
\item[$i.)$] 
if at least one coordinates $u_j = 0$,
then $C(u) = 0$;
\item[$ii.)$] $C$ is $d$-increasing, \emph{i.e.},  for every $a = (a_1, \ldots, a_d)$ and $b= (b_1, \ldots, b_d)$ 
in $[0,1]^d$ such that $a_i \le b_i$, $i=1, \ldots, d$, the $C$-volume
$V_C([a,b])$ of the box $[a,b] = [a_1, b_1] \times \cdot \times [a_d, b_d]$ is positive.
\item[$iii.)$] if $u_j =1$ for all $j \ne k$ for some fixed $k$, 
then $C(u) = u_k$.
\end{itemize}
We can now state Sklar's result. 

\begin{theorem}[Sklar's Theorem]
\label{th:1} 
Every multivariate 
cumulative distribution function (CDF),
	\[
		H (x_1, \ldots , x_d) = P \bigl\{ X_1 \le x_1, \ldots , X_d \le x_d \bigr\} \; ,
	\]
can be expressed in terms of its marginals 
$F_i(x_i) = P \bigl\{X_i \le x_i \bigr\}$, $i = \{1,\ldots, d\}$,  
and a copula $C$, such that 
	\[
		H(x_1, \ldots , x_d) = C \bigl( F_1(x_1), \ldots , F_d(x_d) \bigr) \; .
	\]
\end{theorem}

Using this result, the problem of estimating 
a bivariate distribution function 
$H(s_f (T), q^{-1}(T))$ can be divided into two 
independent problems: 
\begin{itemize}
\item[$i.)$] Estimating the marginal distributions. 
The marginals are the market implied 
cumulative distribution functions of $S_f (T)$ and $Q^{-1}(T)$. 
We  denote them by  $F_{S_f (T)}$ and $F_{Q^{-1}(T)}$; and
\item[$ii.)$] estimating a copula  
	\[
		C = C \Bigl( F_{S_f (T)} \bigl(s_f (T) \bigr) \, , \, 
		F_{Q^{-1}(T)} \bigl( q^{-1}(T) \bigr) \Bigr) \; , 
	\] 
which specifies the dependence relation between $S_f(T)$ and $Q^{-1}(T)$.   
The existence of such a copula is guaranteed by Sklar's Theorem. 
\color{black} 
\end{itemize}

\noindent
It follows from point~$i.)$ that, as the market 
implied cumulative distribution functions are used, our model 
duly adapts to the observed volatility smile.  

\bigskip
We will address item $i.)$ in Section~\ref{sec:3.2} and item
$ii.)$ in Section~\ref{sec:3.3}. 
\color{black} 

\subsection{The marginals}
\label{sec:3.2}

In order to estimate 
the marginal distributions, the strategy adopted 
by DSW is to calibrate the parameters 
of a single Heston model on the market data of 
plain vanilla option prices, 
for both $S_f$ and $Q^{-1}$. The vectors of parameters for 
each Heston model are denoted by 
$\varphi_{S_f}$ and $\varphi_{Q^{-1}}$, for $S_f$ and $Q^{-1}$,
respectively. 
We will simply take over this first step from DSW 
and consider it as part of our own approach.  

However, for the purpose of illustration only, 
we will use \emph{hypothetical data} 
in Section~\ref{sec:4} and the parameters of the 
$\varphi_{S_f}$ and $\varphi_{Q^{-1}}$ 
vectors will be set directly, \emph{i.e.}, without a calibration 
to real market data. 

\subsection{The copula}
\label{sec:3.3}

According to Theorem~\ref{th:1}, an estimate for
$H \bigl(s_f (T ), q^{-1}(T ) \bigr)$ can be provided once a 
copula $C$ linking the random variables 
$S_f (T)$ and $Q^{-1}(T)$ is identified. 

Our approach is  
to calibrate the copula $C$ using data 
provided by an expert. The data are represented by a $(N \times 2)$ 
matrix $\mathbb{A}$, the first column contains data of $S_f (T)$, and 
the second column contains data of $Q^{-1}(T)$. By $\mathbb{A}(n)$, 
$n = \{1, \ldots , N \}$, we denote the $n$-th line of $\mathbb{A}$. 
$N$ is the number of ordered pairs provided by the expert.

In order to build a copula based on the matrix~$\mathbb{A}$, we make use of 
kernel estimators\footnote{We refer to \cite{Gramacki}
for the theory of kernel density estimation.}, 
following the methodology proposed by Scaillet and 
Fermanian \cite[Section~3.1]{FS}. 
The role of the kernels is to smoothen the data. 
In case there are sufficient data, the 
obtained bivariate CDF does not 
depend on the choice of a particular Kernel estimator. Hence, we work 
with $d$-dimensional Gaussian Kernel functions of the form
	\[
		K ( x )= (2\pi)^{-\frac{d}{2}} \; {\rm e}^{- \frac{1}{2}
		x^T x}  \; , 
		\qquad  x= (x_1, \ldots, x_d) \; . 
	\]
As one may expect, the probability density function related to 
our empirical CDF 
places more probability mass where there are 
more ordered pairs, and less probability mass where 
there are less ordered pairs. 

%

The estimated bivariate cumulative 
distribution function (CDF) of the two dependent 
random variables $S_f$ and $Q^{-1}$, 
denoted by $\widehat{F}$, is given by
	\begin{align*}
		\widehat{F} (s_f,q^{-1}) 
		& =
		\int_{- \infty}^{s_f}  {\rm d} s \int_{- \infty}^{q^{-1}}  {\rm d} r \; \; 
		\widehat{f} ( s , r) \;  ,
	\end{align*}
with $\widehat{f}( s , r) 
		  = \frac{1}{N h^2 }\sum_{n=1}^N 
		K \left( \tfrac{( s , r) -  \mathbb{A}(n)}{h} \right)   
	$
the Kernel estimator of $f(s,r)$. 

\bigskip
We are now able to define the copula which 
will allow us to compute the price of a quanto option. 

\begin{definition}
A copula~$C$ is obtained by setting
	\begin{equation}
	\label{(8)}
		C(u_1, u_2) \equiv \widehat{F} 
		\bigl( \xi_1(u_1), \xi_2(u_2) \bigr) \; ,  
	\end{equation}
where $\xi_1 (u_1) 
= \inf \, \bigl\{ y_1 \mid \widehat{F}_{S_f} (y_1) \ge u_1 \bigr\}$ 
and  $\xi_2 (u_2) 
= \inf \, \bigl\{ y_2 \mid \widehat{F}_{Q^{-1}} (y_2) \ge u_2 \bigr\}$. 
\end{definition}

\begin{remark}
One easily verifies that 
the greater the number $N$ of ordered 
pairs provided by the expert, the lower the impact of the 
choice of the kernel function $K$ and the 
bandwidth~$h$, on the copula estimation. 
\end{remark}

We now state the 
relation between 
$\rho(S_f, Q)$ (the correlation between the infinitesimal increments of 
$S_f$ and $Q$) and $\rho(S_f, Q^{-1})$ (the correlation between 
the infinitesimal increments of 
$S_f$ and $Q^{-1}$).  
This information will be used in the numerical 
illustration section, in order to allow the three 
approaches to be compared, as the practitioners 
approach is based on the relation between
$S_f$ and $Q$, while the 
DSW approach and our approach are based on the 
relation between $S_f$ and $Q^{-1}$. 

\begin{proposition}
\label{prop:3.5}
$\rho(S_f, Q) = - \rho(S_f, Q^{-1})$. 
\end{proposition}
\color{black}

\begin{proof}
Without loss of generality, only stochastic terms shall be 
considered. From \eqref{(4)}, it follows that
	\[
		{\rm d} Q(t) = Q(t) \sqrt{V_2} \, {\rm d} W_{ \mathbb{Q}_{2}}(t) \; .
	\]
The difference between the $Q(t)$ diffusion, under the 
domestic and the foreign risk-neutral measure, lies 
in the drift term. The format of the Brownian motion part 
remains unaltered. Thus, under the foreign risk-neutral 
measure~$W_{ \mathbb{Q}_{f_2}}$, 
	\[
		 {\rm d}Q(t) = Q(t) \sqrt{V_2} \, {\rm d} W_{ \mathbb{Q}_{f_2}} (t) \; . 
	\]
We apply It\o's Lemma to $Q^{-1}$. We find
	\[
   		{\rm d} Q^{-1}(t) =
		Q^{-1} (t) \left(- \sqrt{V_2} \right) \, {\rm d} W_{ \mathbb{Q}_{f_2}} (t).
	\]
Inspecting equation \eqref{(5)}, 
we get, under the foreign risk neutral measure, 
	\[
		{\rm d} Q^{-1}(t) = Q^{-1} (t)  \sqrt{V_2} \left(
		- \rho (S_f, Q) {\rm d} W_{\mathbb{Q}_{f_1}} - 
		\sqrt{ 1 - \rho^2 (S_f, Q)} \; {\rm d} W_{\mathbb{Q}_{f_3}} \right) \; ,  
	\]
where $W_{\mathbb{Q}_{f_1}}$ and $W_{\mathbb{Q}_{f_3}}$ are 
independent Brownian motions. Under the foreign risk-neutral 
measure, the stochastic process $S_f$ satisfies 
	\begin{align*}
		{\rm d} S_f (t) = r_f S_f (t) {\rm d} t
		+ \sqrt{V_1(t)} \,  S_f (t) {\rm d} W_{\mathbb{Q}_{f_1}} (t) \; . 
	\end{align*}
Hence,
	\begin{align*}
		\rho(S_f, Q^{-1})  
		&  = \operatorname{Cor} \; 
		\left[ {\rm d} W_{\mathbb{Q}_{f_1}}(t), 
		\left(
		- \rho (S_f, Q) {\rm d} W_{\mathbb{Q}_{f_1}} - 
		\sqrt{ 1 - \rho^2 (S_f, Q)} \; {\rm d} W_{\mathbb{Q}_{f_3}} \right) 
		\right]
		\\
		& = - \operatorname{Cor} \; \left[ {\rm d} W_{\mathbb{Q}_{f_1}}(t), 
		\left(
		 \rho (S_f, Q) {\rm d} W_{\mathbb{Q}_{f_1}} + 
		\sqrt{ 1 - \rho^2 (S_f, Q)} \; {\rm d} W_{\mathbb{Q}_{f_3}} \right) 
		\right] \; ; 
	\end{align*}
thus $ \rho(S_f, Q^{-1}) = - \rho(S_f, Q)$.
\end{proof}

\section{Numerical illustration}
\label{sec:4}

In order to analyse the pricing differences 
among the practitioners' framework, the DSW 
framework, and our approach based on empirical copulas,
we proceed as follows: we set numerical values displayed 
in the following table. They are used in \emph{all} the cases 
we will discuss.

\begin{table}[h!]
  \begin{center}
    \vskip .1cm
    \label{tab:table1}
    \begin{tabular}{r|r|r|r|r|r}  
      \textbf{correlation}  & \textbf{initial}
      & \textbf{initial asset} & \textbf{domestic} & \textbf{foreign risk}
      & \textbf{constant}\\
      $\rho(S_f, Q^{-1})$ & \textbf{exchange }
      & \textbf{value} $S_f(0)$ & \textbf{risk free} & \textbf{free interest}
      & \textbf{exchange}\\
         & \textbf{rate} $Q(0)$ &  
      & \textbf{interest} $r$ & 
      \textbf{rate} $r_f$ & \textbf{rate} $q$ \\
       & &  
      & \textbf{rate} $r$ & 
       &  \\
      \hline
      - 0.7  & 3.1 & 2500 & 0.1 & 0.01 & 3
    \end{tabular}
  \end{center}
\end{table}

\goodbreak
\bigskip
\noindent
We will vary 
\begin{itemize}
\item the Heston vector parameters 
$\varphi_{S_f}$ and $\varphi_{Q^{-1}}$; and 
\item the time to maturity~$T$. 
\end{itemize} 
We also set to zero the continuous dividend yield $d(t)$, 
from the DSW approach depicted in Section~\ref{sec:2.2}.  

These choices allow us to 
compute the prices of foreign vanilla call options on both
DOM currency and on $S_f$, 
and to derive the implied volatility smiles of these options:

\begin{itemize} 
\item[$i.)$] We compute the quanto option prices in the 
practitioners' framework, using equation~\eqref{(6)}
and $\rho(S_f, Q) = - \rho(S_f, Q^{-1})$; 
\item[$ii.)$] We  
evaluate the quanto option prices using the DSW framework 
outlined in Section~\ref{sec:2.2};
\item[$iii.)$] We 
compute the proposed quanto option prices in our 
new copula approach: 
\begin{itemize}
\item 
We numerically derive the marginal cumulative distribution functions 
$F_{S_f} \bigl( s_f (T) \bigr)$ and $F_{Q^{-1}} \bigl( q^{-1}(T) \bigr)$, 
respectively, from the Heston model with 
parameters $\varphi_{S_f}$ and $\varphi_{Q^{-1}}$;  

\item  
We compute the copula $C(u_1, u_2)$
form the matrix $\mathbb{A}$ (provided by an external expert),
using equation \eqref{(8)}. Sampling from the copula $C(u_1, u_2)$, 
we obtain ordered 
pairs of quantiles  $(v_1, v_2)$;
 
\item 
The ordered pairs of quantiles 
$(v_1, v_2)$ are 
transformed into $S_f$ and $Q^{-1}$ outcomes, by setting 
	\[
		\Bigl( s_{f} (T) , q^{-1}(T) \Bigr) 
		= \left( F^{-1}_{S_f} (v_1) , F_{Q^{-1}}^{-1} (v_2) \right)  \; . 
	\]
\item  
For each obtained ordered pair $\bigl( s_f (T ), q^{-1}(T ) \bigr)$,  
equation \eqref{(7)} yields 
	\[
		C^q (0) = Q(0) {\rm e}^{- r_f T} q^{-1}( T ) 
		\max \bigl\{ q (s_f (T ) - K ), 0 \bigr\} \; .
	\]

\end{itemize}
The \emph{average} of the numerous obtained 
values of $C^q(0)$ is  the price 
we propose  for of the quanto option in 
the empirical copula dependence relation framework.
\end{itemize}

\bigskip
We now discuss the outcome of these three procedures 
for different volatility smiles and 
dependence relation fashions between $S_f$ and $Q^{-1}$,
in a case by case analysis.

\subsection{Case I: Gaussian copula, 
constant volatility}

The matrix $\mathbb{A}$ is set such that the 
obtained copula $C$ is Gaussian with correlation 
$\rho(S_f, Q^{-1})$ and the parameters 
$\varphi_{S_f}$ and $\varphi_{Q^{-1}}$ are set to 
	\[
		\varphi_{S_f}=\varphi_{Q^{-1}} = (0, 0,0, 0.2, 0)
	\] 
(whence no volatility smile is present for both $S_f$ and $Q^{-1}$). 
The time to maturity is $T=3$. 

The DSW \cite{DSW} and the
empirical copula approaches are capable of adapting to 
the imposed constant volatility 
smile, as these approaches can even adapt to non-constant 
volatility smiles.
Since the  
$S_f$ and $Q^{-1}$ diffusions 
are correlated by a simple
constant correlation $\rho(S_f,Q^{-1})$, 
the basic 
assumption of the practitioners' framework is satisfied. 
The DSW framework   
and the empirical copula approach 
are capable of adapting to this condition as well: 
the DSW model directly uses $\rho(S_f,Q^{-1})$ to 
correlate 
$S_f$ and $Q^{-1}$
diffusions,
and the empirical copula approach 
simply reproduces the Gaussian copula dependence relation with 
correlation $\rho(S_f, Q^{-1})$ from the data given 
by the matrix $\mathbb{A}$. Hence, \emph{no pricing differences 
are observed} amongst the three approaches, 
despite minor differences due to simulation imprecisions.

\subsection{Case II: Gaussian copula, co-inclining volatility smile}
 
\begin{figure}[h]
\includegraphics[width=0.52\textwidth]{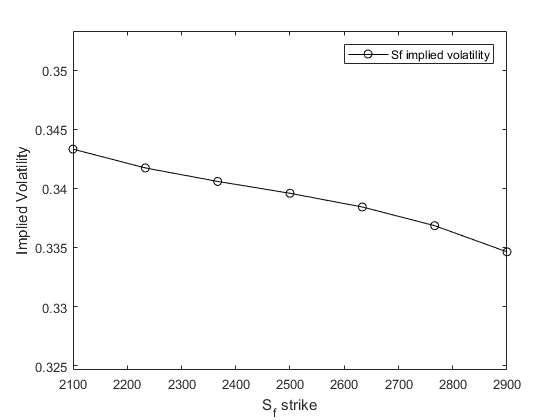}
\includegraphics[width=0.52\textwidth]{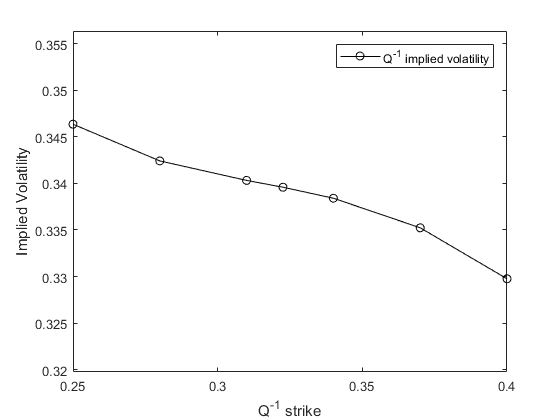}
\caption{Implied volatilities for $S_f$ and $Q^{-1}$, 
as a function of strike, subject to a   
co-inclining volatility smile.}
\label{fig1}
\end{figure}

The matrix $\mathbb{A}$ is set such that the obtained 
copula is Gaussian, and 
	\[ 
		\varphi_{S_f} = \varphi_{Q^{-1}}  
		= \bigl( - 0.7 , 1, 0.1, 0.2, 0.5 \bigr) \; , 
	\] 
whence a co-inclining volatility smile is obtained for
$S_f$ and $Q^{-1}$ (as can be seen from their 
vectors of parameters $\varphi_{S_f}$ and $\varphi_{Q^{-1}}$, 
and Figure~\ref{fig1}). The time to maturity is $T=3$. 

\begin{figure}[h]
\includegraphics[width=12cm]{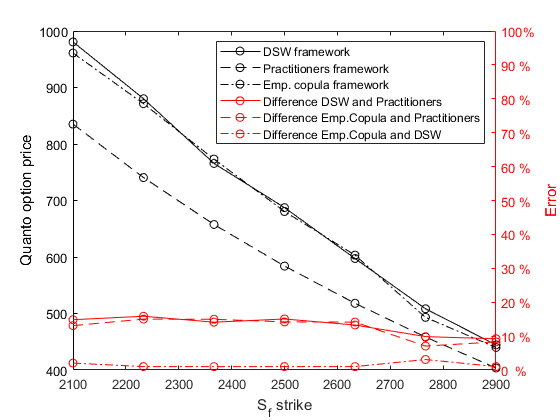}
\caption{Pricing differences, Gaussian copula, co-inclining volatility smile.}
\label{fig3}
\end{figure}

Both the DSW approach and the empirical copula 
approach adapt to  
the volatility smiles, while the practitioners' 
approach does not, because of its ``volatility is constant'' assumption.
Concerning the dependence relation between $S_f$ and $Q^{-1}$, 
the analysis is the same as in 
case I. Hence, no pricing differences should be observed 
between the DSW and the empirical copula 
frameworks (see Fig.~\ref{fig3}). The minor 
differences observed between these two approaches  
are due to simulation imprecisions.

\subsection{Case III: $t$-student copula, long term option}

In Cases III and IV, 
the matrix $\mathbb{A}$ is set such that the obtained 
copula is a $t$-copula 
with $3$ degrees of freedom and correlation $\rho(S_f, Q^{-1})$, 
	\[
		\varphi_{S_f} = \varphi_{Q^{-1}} = ( - 0.7, 1,0 . 1, 0.2, 0.5)
	\] 
(whence a co-inclining volatility smile is obtained for
$S_f$ and $Q^{-1}$, which is displayed in 
Figure~\ref{fig1}). The time to maturity is $T=3$. 

\begin{figure}[h]
\includegraphics[width=12cm]{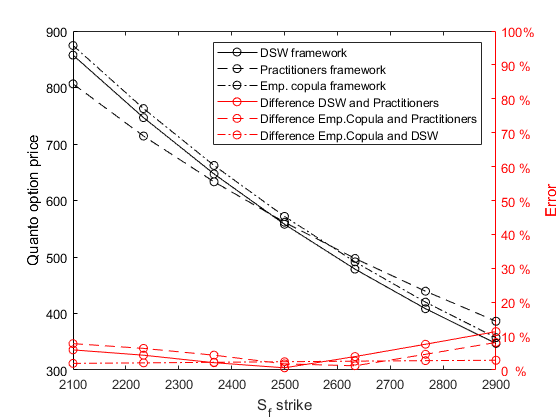}
\caption{Pricing differences, $t$-student copula, long term option.}
\label{fig4}
\end{figure}

The practitioners' framework is not capable of adapting 
to this case, because of the imposed volatility smile; 
and the DSW framework is not capable to 
adapt to this case either: while it is capable to adapt to the 
volatility smile, it is \emph{not able to adapt} to the $t$-copula between 
$S_f$ and $Q^{-1}$. The latter  
presents more tail dependence than 
the Gaussian copula, which is intrinsic to the
DSW framework. Hence, pricing differences 
are observed amongst all the three frameworks 
(see Figure~\ref{fig4}). 
The slight difference between the DSW 
framework and our framework is attributed 
to the difference between a $t$-copula (with 3~degrees of freedom) 
and a Gaussian copula, with the same correlation parameter.

\vfill

\subsection{Case IV: $t$-student copula, short term option}

In Case IV, the conditions are exactly 
the same as in Case III, except that $T = 0.25$ instead of $T = 3$.

\newpage
\begin{figure}[h!]
\includegraphics[width=12cm]{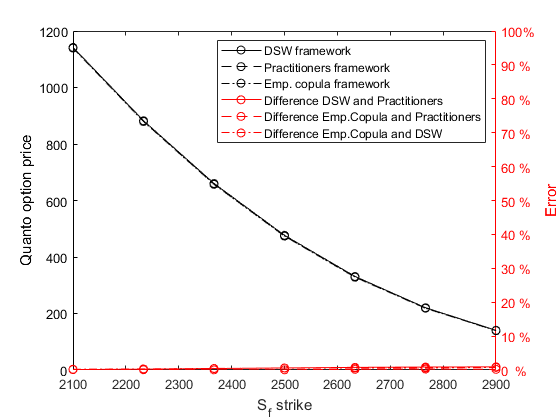}
\caption{Pricing differences, $t$-student copula, short term option.}
\label{fig5}
\end{figure}

Figure~\ref{fig5} shows that  
\emph{no pricing differences are observed}. 
We conclude that neither the 
dependence relation between $S_f$ and $Q^{-1}$ 
nor the volatility smile play a major role in the pricing
of quanto options, if the contract is a short-term call option.

\subsection{Case V: Frank copula, long term option}

In this case, the same simple conditions as in 
Case $I$ are imposed, except that~$\mathbb{A}$ is set such 
that the obtained 
copula is a Frank copula with parameter $\alpha$.  The 
ordered 
pairs of quantiles generated by this copula, when converted 
to ordered pairs of normal random variables, induce a correlation
$\rho (S_f, Q^{-1})$. 
A Frank copula is less similar to a Gaussian copula than a $t$-copula is. 
Whence, this case stresses the 
modelling of the dependence relation 
more than Case $III$ does.    

\begin{figure}[h!]
\begin{center}
\includegraphics[width=12cm]{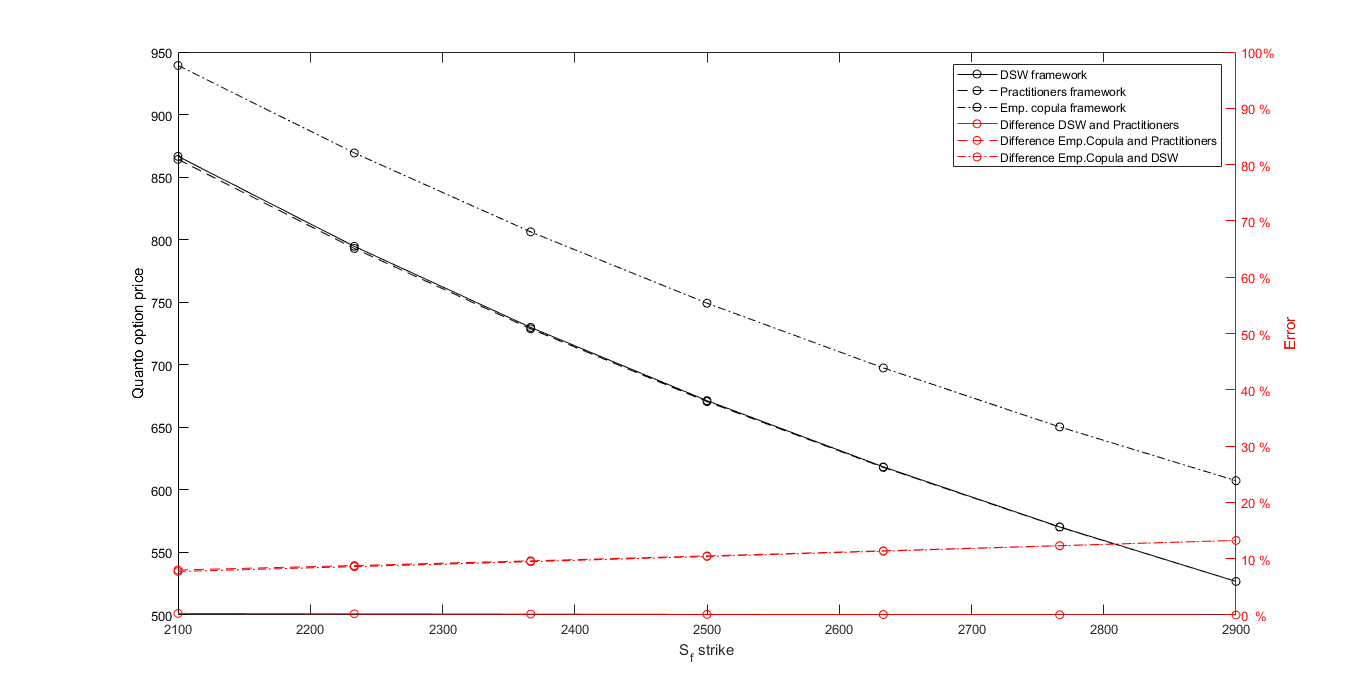}
\caption{Pricing differences, Frank copula, long term option.}
\label{fig6}
\end{center}
\end{figure}

The DSW and the practitioners' frameworks yield similar results, 
as no volatility smile is imposed and both approaches adapt to 
the imposed Frank copula dependence relation the same way, 
\emph{i.e.},  
by considering solely its induced correlation.  The empirical copula 
framework gives pricing figures considerably different from the 
other approaches as it takes into account the full dependence 
relation provided by the imposed Frank copula. 
Figure~\ref{fig6} illustrates these results.

\subsection{Case VI: Frank copula, short term option}

The conditions are the same as in Case V, except that now $T= 
0,25$.
As a consequence, major pricing differences 
among the three models are not identified, even though slight 
pricing differences   
for deep out-of-the money options exist. 

\begin{figure}[h]
\includegraphics[width=12cm]{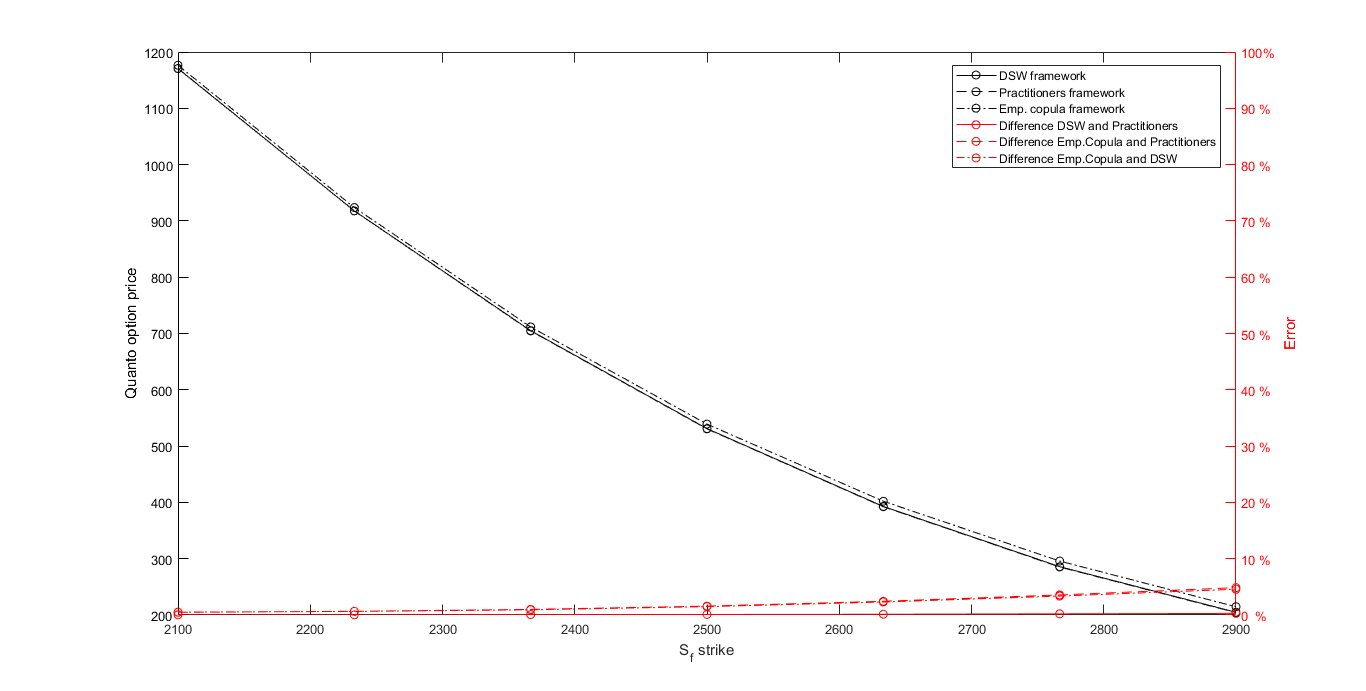}
\caption{Pricing differences, Frank copula, short term option}
\label{fig7}
\end{figure}

Figure~\ref{fig7} 
illustrates the pricing differences. Whence, even 
in a stressed dependence relation context, the dependence 
relation does not play a major role in the pricing of short-term 
quanto options.

\section{Summary}

We have proposed a framework based on empirical copulas 
for quanto option pricing. We have given numerical examples 
in order to illustrate the pricing differences among our 
approach and the practitioners as well as the DSW model \cite{DSW}. 
Looking at the results, we conclude that: 
\begin{itemize}
\item[$i.)$]  
the quanto option requires 
explicit modelling for accurate pricing, with the exception 
of short duration contracts; 
\item[$ii.)$] the flexibility provided 
by the empirical copula approach results in pricing differences 
when compared to the other two approaches. 
\end{itemize}

On the proposed 
empirical copula dependence relation framework, we 
conclude that: 
\begin{itemize}
\item[$iii.)$] it provides a flexible framework to define 
the dependence relation between the market variables used 
in quanto option pricing, by taking into account non-linear 
dependence relations, 
through the matrix 
$\mathbb{A}$ and the related empirical copula 
estimation framework; 
\item[$iv.)$] it can adapt to 
the observed volatility smiles from the relevant market 
variables, as the marginals of $S_f$ and $Q^{-1}$ shall be
calibrated based on plain vanilla options market prices; and finally
\item[$v.)$] a 
drawback of the proposed model is that it is computationally 
more expensive than the other models it was compared to.
\end{itemize}

\end{document}